\documentclass[12pt]{article}
\usepackage{epsf}
\usepackage{rotate}
\renewcommand{\thefootnote}{\fnsymbol{footnote}}

\begin{document}

%%%%%%%%%%% Titlepage %%%%%%%%%%% 

\begin{titlepage}
\begin{flushright}
\begin{tabular}{l}
DESY 01--006\\
CERN--TH/2001--012\\
TTP01--06\\
hep--ph/0101276\\
January 2001
\end{tabular}
\end{flushright}

\vspace*{1.8truecm}

\begin{center}
\boldmath
{\Large \bf General Analysis of New Physics in $B\to J/\psi K$}
\unboldmath

\vspace*{2.1cm}

\smallskip
\begin{center}
{\sc {\large Robert Fleischer}}\footnote{E-mail: {\tt 
Robert.Fleischer@desy.de}} \\
\vspace*{2mm}
{\sl Deutsches Elektronen-Synchrotron DESY, Notkestra\ss e 85, 
D--22607 Hamburg, Germany}
\vspace*{1truecm}\\
{\sc{\large Thomas Mannel}}\footnote{E-mail: {\tt
              Thomas.Mannel@cern.ch}}\\
\vspace*{2mm}
{\sl CERN Theory Division, CH--1211 Geneva 23, Switzerland}
\\ and \\
{\sl Institut f\"{u}r Theoretische Teilchenphysik,
     Universit\"{a}t Karlsruhe, \\ D--76128 Karlsruhe, Germany}
\end{center}

\vspace{2.0truecm}

{\large\bf Abstract\\[10pt]} \parbox[t]{\textwidth}{
We present a model-independent parametrization of the 
$B^\pm\to J/\psi K^\pm$, $B_d\to J/\psi K_{\rm S}$ decay amplitudes by 
taking into account the constraints that are implied by the isospin 
symmetry of strong interactions. Employing estimates borrowed from 
effective field theory, we explore the impact of physics beyond the 
Standard Model and introduce -- in addition to the usual mixing-induced 
CP asymmetry ${\cal A}_{\rm CP}^{\rm mix}$ in $B_d\to J/\psi K_{\rm S}$ -- 
a set of three observables, allowing a general analysis of possible 
new-physics effects in the $B\to J/\psi K$ system. Imposing a dynamical 
hierarchy of amplitudes, we argue that one of these observables may 
already be accessible at the first-generation $B$-factories, whereas 
the remaining ones will probably be left for second-generation $B$ 
experiments. However, in the presence of large rescattering effects, all 
three new-physics observables may be sizeable. We also emphasize that  
a small 
value of ${\cal A}_{\rm CP}^{\rm mix}$ could be due to new-physics effects 
arising at the $B\to J/\psi K$ decay-amplitude level. In order to establish 
such a scenario, the observables introduced in this paper play a key role.}

\vskip1.5cm

\end{center}

\end{titlepage}

\thispagestyle{empty}
\vbox{}
\newpage
 
\setcounter{page}{1}

\setcounter{footnote}{0}
\renewcommand{\thefootnote}{\arabic{footnote}}

\section{Introduction}\label{sec:intro}
At present, we are at the beginning of the $B$-factory era in particle
physics, which will provide valuable insights into CP violation and
various tests of the Kobayashi--Maskawa picture of this phenomenon 
\cite{KM}. Among the most interesting $B$-decay channels is the 
``gold-plated'' mode $B_d\to J/\psi K_{\rm S}$ \cite{bisa}, which allows 
the determination of the angle $\beta$ of the unitarity triangle of
the Cabibbo--Kobayashi--Maskawa (CKM) matrix \cite{revs}. In the summer 
of 2000 -- after important steps at LEP \cite{LEP} and by the CDF 
collaboration \cite{CDF} -- the first results on CP-violating effects 
in $B_d\to J/\psi K_{\rm S}$ were reported by the BaBar \cite{babar} and 
Belle \cite{belle} collaborations, which already led to some excitement 
in the $B$-physics community \cite{sin2b-NP}. 

In this paper, we consider the neutral mode $B_d\to J/\psi K_{\rm S}$ 
together with its charged counterpart $B^\pm\to J/\psi K^\pm$. Making 
use of the isospin symmetry of strong interactions, we derive 
a model-independent parametrization of the corresponding decay amplitudes. 
After a careful analysis of the Standard-Model contributions, we include 
possible new-physics amplitudes and introduce -- in addition to the 
usual ``mixing-induced'' CP asymmetry in $B_d\to J/\psi K_{\rm S}$ -- 
a set of three observables, allowing a general analysis of new-physics 
effects in the $B\to J/\psi K$ system; the generic size of these effects
is estimated with the help of arguments borrowed from effective field
theory. The four observables provided by $B\to J/\psi K$ decays are  
affected by physics beyond the Standard Model in two different ways: through 
$B^0_d$--$\overline{B^0_d}$ mixing, and new-physics contributions to the 
$B\to J/\psi K$ decay amplitudes. Usually, $B^0_d$--$\overline{B^0_d}$ 
mixing is considered as the preferred mechanism for new physics to manifest 
itself in the mixing-induced $B_d\to J/\psi K_{\rm S}$ CP asymmetry 
\cite{NP-revs}. Here we focus on new-physics effects arising at the 
$B\to J/\psi K$ decay-amplitude level \cite{growo}; we emphasize that 
the extraction of the CKM angle $\beta$ from mixing-induced CP violation 
in $B_d\to J/\psi K_{\rm S}$ may be significantly disturbed by such  
effects. As is well known, the value of ``$\beta$'' itself, i.e. the 
CP-violating weak $B^0_d$--$\overline{B^0_d}$ mixing phase, may deviate 
strongly from the Standard-Model expectation because of new-physics 
contributions to $B^0_d$--$\overline{B^0_d}$ mixing. 

The three observables introduced in this paper allow us to search for
``smoking-gun'' signals of new-physics contributions to the $B\to J/\psi K$ 
decay amplitudes, which would also play an important role for  
mixing-induced CP violation in $B_d\to J/\psi K_{\rm S}$. Employing an 
isospin decomposition and imposing a hierarchy of amplitudes, we argue that 
one of these observables may be sizeable and could already be accessible at 
the first-generation $B$-factories. On the other hand, the remaining two 
observables are expected to be dynamically suppressed, but may be within 
reach of the second-generation $B$ experiments, BTeV and LHCb. In the
presence of large rescattering effects, all three observables could
be of the same order of magnitude. However, we do not consider this as 
a very likely scenario and note that also the ``QCD factorization'' 
approach is not in favour of such large rescattering processes \cite{PQCD}. 

The outline of this paper is as follows: in Section~\ref{sec:pheno}, we 
investigate the isospin structure of the $B^\pm\to J/\psi K^\pm$, 
$B_d\to J/\psi K_{\rm S}$ decay amplitudes, and parametrize them in a
model-independent manner; a particular emphasis is given to the 
Standard-Model case. In Section~\ref{sec:NP}, we then have a closer look
at the impact of new physics. To this end, we make use of dimensional 
estimates following from effective field theory, and introduce a plausible 
dynamical hierarchy of amplitudes. The set of observables to search for 
``smoking-gun'' signals of new-physics contributions to the $B\to J/\psi K$ 
decay amplitudes is defined in Section~\ref{sec:obs}, and is discussed in 
further detail in Section~\ref{sec:disc}. In Section~\ref{sec:concl}, our 
main points are summarized.

\boldmath
\section{Phenomenology of $B\to J/\psi K$ Decays}\label{sec:pheno}
\unboldmath
The most general discussion of the $B^\pm\to J/\psi K^\pm$,
$B_d\to J/\psi K_{\rm S}$ system can be performed in terms of an 
isospin decomposition. Here the corresponding initial and 
final states are grouped in the following isodoublets:
\begin{equation}
\left(\begin{array}{c} 
|1/2;+1/2\rangle\\
|1/2;-1/2\rangle\end{array}\right):
\quad
\underbrace{\left(\begin{array}{c} |B^+\rangle  \\ 
|B^0_d \rangle \end{array}\right),
\quad
\left(\begin{array}{c} |\overline{B^0_d}\rangle  \\ 
-|B^- \rangle  
\end{array}\right)}_{{\cal CP}},
\quad
\underbrace{\left(\begin{array}{c} |J/\psi K^+\rangle  \\ 
|J/\psi K^0 \rangle  
\end{array}\right),
\quad
\left(\begin{array}{c} |J/\psi\overline{K^0}\rangle  \\ 
-|J/\psi K^-  \rangle \end{array}\right)}_{{\cal CP}},
\end{equation}
which are related by CP conjugation. As usual, the $K_{\rm S}$ state 
in $B_d\to J/\psi K_{\rm S}$ is the superposition of the two neutral kaon
states $|K^0\rangle$ and $|\overline{K^0}\rangle$ corresponding to CP 
\mbox{eigenvalue $+1$.} The decays $B^+\to J/\psi K^+$ and 
$B^0_d\to J/\psi K^0$ are described by an effective low-energy Hamiltonian 
of the following structure:
\begin{equation}\label{Heff}
{\cal H}_{\rm eff}=\frac{G_{\rm F}}{\sqrt{2}}\left[V_{cs}V_{cb}^\ast
\left({\cal Q}_{\rm CC}^c-{\cal Q}_{\rm QCD}^{\rm pen}-
{\cal Q}_{\rm EW}^{\rm pen}\right)+V_{us}V_{ub}^\ast
\left({\cal Q}_{\rm CC}^u-{\cal Q}_{\rm QCD}^{\rm pen}-
{\cal Q}_{\rm EW}^{\rm pen}\right)\right],
\end{equation}
where the ${\cal Q}$ are linear combinations of perturbatively 
calculable Wilson coefficient functions and four-quark operators,
consisting of current--current (CC), QCD penguin and electroweak (EW) 
penguin operators. For an explicit list of operators and a detailed
discussion of the derivation of (\ref{Heff}), the reader is referred
to \cite{BBL-rev}. For the following considerations, the flavour 
structure of these operators plays a key role:
\begin{equation}\label{CC-def}
{\cal Q}_{\rm CC}^c\sim (\overline{c}c)(\overline{b}s), \quad
{\cal Q}_{\rm CC}^u\sim (\overline{u}u)(\overline{b}s),
\end{equation}
\begin{equation}\label{QCD-def}
{\cal Q}_{\rm QCD}^{\rm pen}\sim \left[(\overline{c}c)+\{(\overline{u}u)
+(\overline{d}d)\}+(\overline{s}s)\right](\overline{b}s),
\end{equation}
\begin{equation}\label{EW-def}
{\cal Q}_{\rm EW}^{\rm pen}\sim \frac{1}{3}\left[2(\overline{c}c)+
\{2(\overline{u}u)-(\overline{d}d)\}-(\overline{s}s)\right](\overline{b}s),
\end{equation}
where the factors of $+2/3$ and $-1/3$ in (\ref{EW-def}) are due to
electrical quark charges. In (\ref{CC-def})--(\ref{EW-def}), we have
suppressed all colour and spin indices. Since
\begin{equation}
(\overline{u}u)=\frac{1}{2}
\underbrace{\left(\overline{u}u+\overline{d}d\right)}_{I=0}+
\frac{1}{2}
\underbrace{\left(\overline{u}u-\overline{d}d\right)}_{I=1},\quad
2(\overline{u}u)-(\overline{d}d)=\frac{1}{2}
\underbrace{\left(\overline{u}u+\overline{d}d\right)}_{I=0}+
\frac{3}{2}
\underbrace{\left(\overline{u}u-\overline{d}d\right)}_{I=1},
\end{equation}
we conclude that the Hamiltonian (\ref{Heff}) is a combination of 
isospin $I=0$ and $I=1$ pieces:
\begin{equation}\label{ham-decom}
{\cal H}_{\rm eff} = {\cal H}_{\rm eff}^{I=0} + {\cal H}_{\rm eff}^{I=1},
\end{equation}
where ${\cal H}_{\rm eff}^{I=0}$ receives contributions from all of
the operators listed in (\ref{CC-def})--(\ref{EW-def}), whereas 
${\cal H}_{\rm eff}^{I=1}$ is only due to ${\cal Q}_{\rm CC}^u$ and
${\cal Q}_{\rm EW}^{\rm pen}$. Taking into account the isospin flavour
symmetry of strong interactions, we obtain
\begin{eqnarray}
\langle J/\psi K^+|{\cal H}_{\rm eff}^{I=0}|B^+\rangle&=&
+\langle J/\psi K^0|{\cal H}_{\rm eff}^{I=0}|B^0_d\rangle\\
\langle J/\psi K^+|{\cal H}_{\rm eff}^{I=1}|B^+\rangle&=&
-\langle J/\psi K^0|{\cal H}_{\rm eff}^{I=1}|B^0_d\rangle,
\end{eqnarray}
and finally arrive at 
\begin{equation}\label{AMPLp1}
A(B^+\to J/\psi K^+)=\frac{G_{\rm  
F}}{\sqrt{2}}\left[V_{cs}V_{cb}^\ast\left\{
A_c^{(0)}-A_c^{(1)}\right\}+V_{us}V_{ub}^\ast\left\{A_u^{(0)}-
A_u^{(1)}\right\}\right]
\end{equation}
\begin{equation}\label{AMPLd1}
A(B^0_d\to J/\psi K^0)=\frac{G_{\rm F}}{\sqrt{2}}\left[V_{cs}V_{cb}^\ast
\left\{A_c^{(0)}+A_c^{(1)}\right\}+V_{us}V_{ub}^\ast
\left\{A_u^{(0)}+A_u^{(1)}\right\}\right],
\end{equation}
where 
\begin{equation}\label{ampl-c}
A_c^{(0)}=A_{\rm CC}^c-A_{\rm QCD}^{\rm pen}-A_{\rm EW}^{(0)},\quad
A_c^{(1)}=-A_{\rm EW}^{(1)}
\end{equation}
\begin{equation}
A_u^{(0)}=A_{\rm CC}^{u (0)}-A_{\rm QCD}^{\rm pen}-A_{\rm EW}^{(0)},\quad
A_u^{(1)}=A_{\rm CC}^{u (1)}-A_{\rm EW}^{(1)}
\end{equation}
denote hadronic matrix elements $\langle J/\psi K|{\cal Q}|B\rangle$, i.e.\
are CP-conserving strong amplitudes. The CKM factors in (\ref{AMPLp1}) and
(\ref{AMPLd1}) are given by
\begin{equation}
V_{cs}V_{cb}^\ast=\left(1-\frac{\lambda^2}{2}\right)\lambda^2A,\quad
V_{us}V_{ub}^\ast=\lambda^4 A\, R_b\, e^{i\gamma},
\end{equation}
with
\begin{equation}
\lambda\equiv|V_{us}|=0.22, \quad A\equiv|V_{cb}|/\lambda^2=0.81\pm0.06, 
\quad R_b\equiv|V_{ub}/(\lambda V_{cb})|=0.41\pm0.07,
\end{equation}
and $\gamma$ is the usual angle of the unitarity triangle of the CKM 
matrix \cite{revs}. Consequently, we may write
\begin{equation}\label{AMPLp2}
A(B^+\to J/\psi K^+)=\frac{G_{\rm F}}{\sqrt{2}}
\left(1-\frac{\lambda^2}{2}\right)\lambda^2A\left\{A_c^{(0)}-A_c^{(1)}
\right\}\left[1+\frac{\lambda^2 R_b}{1-\lambda^2/2}
\left\{\frac{A_u^{(0)}-A_u^{(1)}}{A_c^{(0)}-A_c^{(1)}}\right\}
e^{i\gamma}\right]
\end{equation}
\begin{equation}\label{AMPLd2}
A(B^0_d\to J/\psi K^0)=\frac{G_{\rm F}}{\sqrt{2}}
\left(1-\frac{\lambda^2}{2}\right)\lambda^2A\left\{A_c^{(0)}+A_c^{(1)}
\right\}\left[1+\frac{\lambda^2 R_b}{1-\lambda^2/2}
\left\{\frac{A_u^{(0)}+A_u^{(1)}}{A_c^{(0)}+A_c^{(1)}}\right\}
e^{i\gamma}\right].
\end{equation}

Let us note that (\ref{AMPLd2}) takes the same form as the parametrization
derived in \cite{RF-BdspsiK}, making, however, its isospin decomposition 
explicit. An important observation is that the CP-violating phase factor
$e^{i\gamma}$ enters in (\ref{AMPLp2}) and (\ref{AMPLd2}) in a doubly 
Cabibbo-suppressed way. Moreover, the $A_u^{(0,1)}$ amplitudes are governed 
by penguin-like topologies and annihilation diagrams,
and are hence expected to be suppressed with 
respect to $A_c^{(0)}$, which originates from tree-diagram-like topologies 
\cite{BFM}. In order to keep track of this feature, we introduce a
``generic'' expansion parameter $\overline{\lambda}={\cal O}(0.2)$
\cite{hierarchy}, which is of the same order as the Wolfenstein parameter 
$\lambda=0.22$:
\begin{equation}\label{H1}
\left|A_u^{(0,1)}/A_c^{(0)}\right|={\cal O}(\overline{\lambda}).
\end{equation}
Consequently, the $e^{i\gamma}$ terms in (\ref{AMPLp2}) and (\ref{AMPLd2})
are actually suppressed by ${\cal O}(\overline{\lambda}^3)$. Since 
$A_c^{(1)}$ is due to dynamically suppressed matrix elements of EW penguin 
operators (see (\ref{ampl-c})), we expect
\begin{equation}\label{H2}
\left|A_c^{(1)}/A_c^{(0)}\right|={\cal O}(\overline{\lambda}^3).
\end{equation}
Therefore, we obtain -- up to negligibly small corrections of 
${\cal O}(\overline{\lambda}^3)$ -- the following expression:
\begin{equation}\label{SM-ampl}
A(B^+\to J/\psi K^+)=A(B^0_d\to J/\psi K^0)=A_{\rm SM}^{(0)},
\end{equation}
with 
\begin{equation}\label{ASM0}
A_{\rm SM}^{(0)}\equiv 
\frac{G_{\rm F}}{\sqrt{2}}\left(1-\frac{\lambda^2}{2}\right)\lambda^2A\,
A_c^{(0)}.
\end{equation}
Let us note that the plausible hierarchy of strong amplitudes given in 
(\ref{H1}) and (\ref{H2}) may be spoiled by very large rescattering
processes \cite{FSI}. In the worst case, (\ref{SM-ampl}) may receive 
corrections of ${\cal O}(\overline{\lambda}^2)$, and not at the
$\overline{\lambda}^3$ level. However, we do not consider this a very
likely scenario and note that also the ``QCD factorization'' approach 
developed in \cite{PQCD} is not in favour of such large rescattering
effects.

The purpose of this paper is to explore the impact of physics beyond 
the Standard Model on $B \to J/\psi K$ decays. In the next section, we 
investigate the generic size of such effects using dimensional arguments 
following from the framework of effective field theory.

\section{Effects of Physics Beyond the Standard Model}\label{sec:NP}
The generic way of introducing physics beyond the Standard Model is to use 
the language of effective field theory and to write down all possible 
dim-6 operators. Of course, this has been known already for a long time and 
lists of the dim-6 operators involving all the Standard-Model particles 
have been published in the literature \cite{dim-6}. After having introduced 
these additional operators, we construct the generalization of the 
Standard-Model effective Hamiltonian (\ref{Heff}) at the scale 
of the $b$ quark. The relevant operators are again dim-6 operators, the  
coefficients of which now contain a Standard-Model contribution, and a 
possible piece of ``new physics''.

The problem with this point of view is that the list of possible dim-6  
operators contains close to one hundred entries, not yet taking into 
account the flavour structure, making this general approach almost 
useless. However, we are dealing with non-leptonic decays in which 
we are, because of our ignorance of the hadronic matrix elements, 
sensitive  neither to the helicity structure
of the operators nor to their colour 
structure. The only information that is relevant in this case is the 
flavour structure, and hence we introduce the notation
\begin{equation}
\left[(\overline{Q} q_1)(\overline{q}_2 q_3)\right] = \sum\,
[\mbox{Wilson coefficient}]\times
[\mbox{dim-6 operator mediating $Q \to q_1 \overline{q}_2 q_3$}].
\end{equation}
Clearly, this sum is renormalization-group-invariant, and involves, at  
the scale of the $b$ quark, Standard-Model as well as possible 
non-Standard-Model contributions. In particular, it allows us to estimate 
the relative size of a possible new-physics contribution.

\boldmath
\subsection{$B^0_d$--$\overline{B^0_d}$ Mixing: $\Delta B = \pm 2$ 
Operators}\label{sec:mix}
\unboldmath
Using this language, we now consider the $\Delta B = \pm 2$ operators  
relevant to $B^0_d$--$\overline{B^0_d}$ mixing. We have \cite{BF-rev}
\begin{equation}\label{Heff-mix}
{\cal H}_{\rm eff}(\Delta B = + 2) = G \left[(\overline{b}d)(\overline{b}d) 
\right]
\end{equation}
as the relevant flavour structure. Within the Standard Model, 
(\ref{Heff-mix}) originates from the well-known box diagrams, which are 
strongly suppressed by the CKM factor $(V_{td} V_{tb}^\ast)^2$, as
well as by a loop factor
\begin{equation} \label{loop}
\frac{g^2_2}{64 \pi^2} =
\frac{G_{\rm F} M_W^2}{\sqrt{128} \pi^2} \approx 1 \times 10^{-3},
\end{equation}
making the Standard-Model contribution very small, of the 
order of\footnote{Note that we do not write the Inami--Lim function 
coming from the box diagram; this function is of order one and is 
ignored in our estimates. A similar comment applies to perturbative QCD 
corrections.}
\begin{equation}
G_{\rm SM} = \frac{G_{\rm F}}{\sqrt{2}} \left(\frac{G_{\rm F} 
M_W^2}{\sqrt{128} \pi^2}\right)(V_{td} V_{tb}^\ast)^2.
\end{equation}
A new-physics contribution therefore could in principle be as 
large as the Standard-Model piece. If $\Lambda$ is the scale of physics 
beyond the Standard Model, we have
\begin{equation}
G_{\rm NP} = \frac{G_{\rm F}}{\sqrt{2}}\left(\frac{G_{\rm F} 
M_W^2}{\sqrt{128} \pi^2}\right)\frac{M_W^2}{\Lambda^2} e^{-i2\psi},
\end{equation}
where $\psi$ is a possible weak phase, which is induced by the 
new-physics contribution. Finally, we arrive at
\begin{equation}\label{PhiM1}
G = \frac{G_{\rm F}}{\sqrt{2}}\left(\frac{G_{\rm F} 
M_W^2}{\sqrt{128} \pi^2}\right)
\left[(V_{td} V_{tb}^\ast)^2+\frac{M_W^2}{\Lambda^2}  
e^{-i2\psi}\right]\equiv
|R|e^{-i\phi_{\rm M}}.
\end{equation}
The relevant quantity is the (weak) phase $\phi_{\rm M}$ of this  
expression, which enters the ``mixing-induced'' CP-violating asymmetries
\cite{revs}. Using the standard parametrization
\begin{equation}
V_{td} V_{tb}^\ast = \lambda^3 A R_t e^{-i\beta},
\end{equation}
where $R_t\equiv|V_{td}/(\lambda V_{cb})|={\cal O}(1)$, we obtain
\begin{equation}\label{PhiM2}
\tan \phi_{\rm M} =
\frac{\sin(2 \beta) + \varrho^2 \sin (2\psi)}
{\cos (2 \beta) + \varrho^2\cos(2\psi)},
\end{equation}
with
\begin{equation}\label{varrho-def}
\varrho=\left(\frac{1}{\lambda^3 A R_t}\right)\left(\frac{M_W}{\Lambda}
\right).
\end{equation}
Since the factor $\varrho$ can be of order one even for large $\Lambda$, 
there can be a large phase shift in the mixing phase. If we assume, for 
example, $A R_t=1$, this term equals 1 for a new-physics scale of 
$\Lambda\sim 8$\,TeV. Such contributions affect of course not only the 
CP-violating phase $\phi_{\rm M}$, but also the ``strength'' $|R|$ of 
the $B^0_d$--$\overline{B^0_d}$ mixing, which would manifest itself as an 
inconsistency in the usual ``standard analysis'' of the unitarity 
triangle \cite{AL}.

\boldmath
\subsection{Decay Amplitudes: $\Delta B = \pm 1$ Operators}
\unboldmath
We can discuss the operators mediating $\Delta B = \pm 1$ processes on 
the same footing as $B^0_d$--$\overline{B^0_d}$ mixing. 
In the presence of new physics, the corresponding low-energy effective 
Hamiltonian can also be composed in $I=0$ and $I=1$ pieces, as in 
(\ref{ham-decom}). In Section~\ref{sec:pheno}, we had a closer look at 
the Standard-Model contributions, arising from current--current, QCD 
and EW penguin operators. New physics may affect the 
corresponding Wilson coefficients, and may introduce new dim-6 operators, 
modifying (\ref{SM-ampl}) as follows:
\begin{eqnarray}
A(B^+\to J/\psi K^+)&=&A_{\rm SM}^{(0)}
\left[1+\sum_k r_0^{(k)}e^{i\delta_0^{(k)}}e^{i\varphi_0^{(k)}}-
\sum_j r_1^{(j)}e^{i\delta_1^{(j)}}e^{i\varphi_1^{(j)}}
\right]\label{ampl-NPp}\\
A(B^0_d\to J/\psi K^0)&=&A_{\rm SM}^{(0)}
\left[1+\sum_k r_0^{(k)}e^{i\delta_0^{(k)}}e^{i\varphi_0^{(k)}}+
\sum_j  
r_1^{(j)}e^{i\delta_1^{(j)}}e^{i\varphi_1^{(j)}}\right].\label{ampl-NPd} 
\end{eqnarray}
Here $r_0^{(k)}$ and $r_1^{(j)}$ correspond to the $I=0$ and $I=1$ pieces, 
respectively, $\delta_{0}^{(k)}$ and $\delta_{1}^{(j)}$ are CP-conserving 
strong phases, and $\varphi_{0}^{(k)}$ and $\varphi_{1}^{(j)}$ the 
corresponding CP-violating weak phases. The amplitudes for the CP-conjugate 
processes can be obtained straightforwardly from (\ref{ampl-NPp}) and 
(\ref{ampl-NPd}) by reversing the signs of the weak phases. The labels $k$ 
and $j$ distinguish between different new-physics contributions to the  
$I=0$ and $I=1$ sectors.

For the following discussion, we have to make assumptions about the
size of a possible new-physics piece. We shall assume that the new-physics 
contributions to the $I=0$ sector are smaller compared to the  
leading Standard-Model amplitude (\ref{ASM0}) by a factor of order 
$\overline{\lambda}$, i.e.
\begin{equation}\label{gen-strength0}
r_0^{(k)}={\cal O}(\overline{\lambda}).
\end{equation}
In the case where the new-physics effects are even smaller, it is difficult 
to disentangle them from the Standard-Model contribution; this will be 
addressed, together with several other scenarios, in 
Section~\ref{sec:disc}. Parametrizing the new-physics amplitudes again by 
a scale $\Lambda$, we have
\begin{equation}
\frac{G_{\rm F}}{\sqrt{2}}\frac{M_W^2}{\Lambda^2}\sim\overline{\lambda}
\left[\frac{G_{\rm F}}{\sqrt{2}}\,\lambda^2A\right],
\end{equation}
corresponding to $\Lambda\sim 1$\,TeV. Consequently, as in the example 
given after (\ref{varrho-def}), also here we have a generic new-physics 
scale in the TeV regime.

As far as possible new-physics contributions to the $I=1$ sector are 
concerned, we assume a similar ``generic strength'' of the corresponding
operators. However, in comparison with the $I=0$ pieces, the matrix elements 
of the $I=1$ operators, having 
the general flavour structure
\begin{equation}
{\cal Q}\sim (\overline{u}u-\overline{d}d)(\overline{b}s), 
\end{equation}
are expected to suffer from a dynamical suppression. As in (\ref{H1}) 
and (\ref{H2}), we shall assume that this brings another factor of 
$\overline{\lambda}$ into the game, yielding
\begin{equation}\label{gen-strength1}
r_1^{(j)}={\cal O}(\overline{\lambda}^2).
\end{equation}
Employing this kind of counting, the new-physics contributions to the 
$I=1$ sector would be enhanced by a factor of ${\cal O}(\overline{\lambda})$ 
with respect to the $I=1$ Standard-Model pieces. This may actually be the 
case if new physics shows up, for example, in EW penguin processes.

Consequently, we obtain
\begin{equation}\label{ampl-hier1}
A(B\to J/\psi K)=A_{\rm SM}^{(0)}\biggl[1+
\underbrace{{\cal O}(\overline{\lambda})}_{{\rm NP}_{I=0}}+
\underbrace{{\cal O}(\overline{\lambda}^2)}_{{\rm NP}_{I=1}}+
\underbrace{{\cal O}(\overline{\lambda}^3)}_{{\rm SM}}\biggr].
\end{equation}
In the presence of large rescattering effects, the assumed dynamical 
suppression through a factor of ${\cal O}(\overline{\lambda})$ would no 
longer be effective, thereby modifying (\ref{ampl-hier1}) as follows:
\begin{equation}\label{ampl-hier2}
\left.A(B\to J/\psi K)\right|_{\rm res.}=A_{\rm SM}^{(0)}\biggl[1+
\underbrace{{\cal O}(\overline{\lambda})}_{{\rm NP}_{I=0}}+
\underbrace{{\cal O}(\overline{\lambda})}_{{\rm NP}_{I=1}}+
\underbrace{{\cal O}(\overline{\lambda}^2)}_{{\rm SM}}\biggr].
\end{equation}
However, as we have noted above, we do not consider this as a very
likely scenario, and shall use (\ref{ampl-hier1}) in the following 
discussion, neglecting the Standard-Model pieces of 
${\cal O}(\overline{\lambda}^3)$, which are not under theoretical
control.

Concerning the analysis of CP violation, it is obvious
that possible weak phases 
appearing in the new-physics contributions play the key role. As was 
the case for the $\Delta B=\pm2$ operators, also the $\Delta B = \pm 1$ 
operators could carry such new weak phases, which would then affect the 
CP-violating $B\to J/\psi K$ observables.

\boldmath
\section{Observables of $B\to J/\psi K$ Decays}\label{sec:obs}
\unboldmath
The decays $B^+\to J/\psi K^+$, $B^0_d\to J/\psi K^0$ and their
charge conjugates provide a set of four decay amplitudes $A_i$.
Measuring the corresponding rates, we may determine the $|A_i|^2$.
Since we are not interested in the overall normalization of the
decay amplitudes, we may construct three independent observables
with the help of the $|A_i|^2$:
\begin{equation}\label{def-AP}
{\cal A}_{\rm CP}^{(+)}\equiv
\frac{|A(B^+\to J/\psi K^+)|^2-|A(B^-\to J/\psi K^-)|^2}{|A(B^+\to 
J/\psi K^+)|^2+|A(B^-\to J/\psi K^-)|^2}
\end{equation}
\begin{equation}\label{def-A0}
{\cal A}_{\rm CP}^{{\rm dir}}\equiv
\frac{|A(B^0_d\to J/\psi K^0)|^2-|A(\overline{B^0_d}\to J/\psi
\overline{K^0})|^2}{|A(B^0_d\to J/\psi K^0)|^2+
|A(\overline{B^0_d}\to J/\psi\overline{K^0})|^2}
\end{equation}
\begin{equation}\label{def-A}
B \equiv\frac{\langle|A(B_d\to J/\psi K)|^2\rangle-
\langle|A(B^\pm\to J/\psi K^\pm)|^2\rangle}{\langle|A(B_d\to 
J/\psi K)|^2\rangle+\langle|A(B^\pm\to J/\psi K^\pm)|^2\rangle},
\end{equation}
where the ``CP-averaged'' amplitudes are defined as follows:
\begin{eqnarray}
\langle|A(B_d\to J/\psi K)|^2\rangle&\equiv&
\frac{1}{2}\left[|A(B^0_d\to J/\psi K^0)|^2
+|A(\overline{B^0_d}\to J/\psi \overline{K^0})|^2\right]\\
\langle|A(B^\pm\to J/\psi K^\pm)|^2\rangle&\equiv&
\frac{1}{2}\left[|A(B^+\to J/\psi K^+)|^2+|A(B^-\to J/\psi K^-)|^2\right].
\end{eqnarray}
If we consider the neutral decays $B_d\to J/\psi K_{\rm S}$, where the
final state is a CP eigenstate with eigenvalue $-1$, the following
time-dependent CP asymmetry provides an additional observable, which
is due to interference between $B^0_d$--$\overline{B^0_d}$ mixing and
decay processes \cite{revs}:
\begin{equation}\label{time-dep}
\frac{\Gamma(B^0_d(t)\to J/\psi K_{\rm S})-
\Gamma(\overline{B^0_d}(t)\to J/\psi K_{\rm S})}{\Gamma(B^0_d(t)\to 
J/\psi K_{\rm S})+\Gamma(\overline{B^0_d}(t)\to J/\psi K_{\rm S})}
={\cal A}_{\rm CP}^{{\rm dir}}\,\cos(\Delta M_d t)+
{\cal A}_{\rm CP}^{\rm mix}\,\sin(\Delta M_d t).
\end{equation}
Here the rates correspond to decays of initially, i.e.\ at time $t=0$,
present $B^0_d$- or $\overline{B^0_d}$-mesons, and $\Delta M_d$
denotes the mass difference between the $B_d$ mass eigenstates. The
``direct'' CP-violating contribution ${\cal A}_{\rm CP}^{{\rm dir}}$ was
already introduced in (\ref{def-A0}), and the ``mixing-induced''
CP asymmetry is given by
\begin{equation}\label{def-mix}
{\cal A}^{\mbox{{\scriptsize mix}}}_{\mbox{{\scriptsize CP}}}
=\frac{2\,\mbox{Im}\,\xi}{1+|\xi|^2}\,,
\end{equation}
with
\begin{equation}\label{xi-def}
\xi=e^{-i\phi}\left[\frac{1+\sum_k r_0^{(k)}e^{i\delta_0^{(k)}}
e^{-i\varphi_0^{(k)}}+\sum_j r_1^{(j)}e^{i\delta_1^{(j)}}
e^{-i\varphi_1^{(j)}}}{1+\sum_k r_0^{(k)}e^{i\delta_0^{(k)}}
e^{+i\varphi_0^{(k)}}+\sum_j r_1^{(j)}e^{i\delta_1^{(j)}}
e^{+i\varphi_1^{(j)}}}\right].
\end{equation}
In (\ref{xi-def}), we have used the parametrization (\ref{ampl-NPd})
to express the corresponding decay amplitudes. The CP-violating weak
phase $\phi$ is given by $\phi=\phi_{\rm M}+\phi_{K}$, where 
$\phi_{\rm M}$ was introduced in (\ref{PhiM1}), and $\phi_K$
is the weak $K^0$--$\overline{K^0}$ mixing phase, which is negligibly
small in the Standard Model. Owing  to the small value of the CP-violating
parameter $\varepsilon_K$ of the neutral kaon system, $\phi_K$ can only
be affected by very contrived models of new physics \cite{NS}.

In order to search for new-physics effects in the $B\to J/\psi K$ system, 
it is useful to introduce the following combinations of the observables 
(\ref{def-AP}) and (\ref{def-A0}):
\begin{equation}
S\equiv\frac{1}{2}\left[{\cal A}_{\rm CP}^{{\rm dir}}+{\cal A}_{\rm  
CP}^{(+)}
\right],\quad
D\equiv\frac{1}{2}\left[{\cal A}_{\rm CP}^{{\rm dir}}-{\cal A}_{\rm  
CP}^{(+)}
\right].
\end{equation}
Using the parametrizations (\ref{ampl-NPp}) and (\ref{ampl-NPd}), and
assuming the hierarchy in (\ref{ampl-hier1}), we obtain
\begin{equation}\label{S-expr}
S=-2\left[\sum_k r_0^{(k)}\sin\delta_0^{(k)}\sin\varphi_0^{(k)}\right]
\left[1-2\sum_l r_0^{(l)}\cos\delta_0^{(l)}\cos\varphi_0^{(l)}\right]
={\cal O}(\overline{\lambda}) + {\cal O}(\overline{\lambda}^2)
\end{equation}
\begin{equation}\label{D-expr}
D=-2\sum_j r_1^{(j)}\sin\delta_1^{(j)}\sin\varphi_1^{(j)}=
{\cal O}(\overline{\lambda}^2)
\end{equation}
\begin{equation}\label{B-expr}
B=+2\sum_j r_1^{(j)}\cos\delta_1^{(j)}\cos\varphi_1^{(j)}=
{\cal O}(\overline{\lambda}^2),
\end{equation}
where terms of ${\cal O}(\overline{\lambda}^3)$, including also a
Standard-Model contribution, which is not under theoretical control,
have been neglected. Note that if the dynamical suppression of the
$I=1$ contributions would be larger, $B$ and $D$ would be further
suppressed relative to $S$.

The corresponding expression for the mixing-induced 
CP asymmetry (\ref{def-mix}) is rather complicated and not very  
instructive. Let us give it for the special case where the new-physics 
contributions to the $I=0$ and $I=1$ sectors involve either the same 
weak or strong phases:
\begin{eqnarray}\label{Amix-calc}
{\cal A}^{\mbox{{\scriptsize mix}}}_{\mbox{{\scriptsize CP}}}&=&
-\sin\phi-2\,r_0\cos\delta_0\sin\varphi_0\cos\phi-
2\,r_1\cos\delta_1\sin\varphi_1\cos\phi\nonumber\\
&&+\,r_0^2\Bigl[\left(1-\cos2\varphi_0\right)\sin\phi+
\cos2\delta_0\sin2\varphi_0\cos\phi\Bigr]=-\sin\phi+
{\cal O}(\overline{\lambda}) + {\cal O}(\overline{\lambda}^2).
\end{eqnarray}
Expressions (\ref{S-expr})--(\ref{B-expr}) also simplify in this case:
\begin{equation}
S=-2\,r_0\sin\delta_0\sin\varphi_0+r_0^2\sin2\delta_0\sin2\varphi_0,\,
D=-2\,r_1\sin\delta_1\sin\varphi_1,\,
B=2\,r_1\cos\delta_1\cos\varphi_1.
\end{equation}
In the following section, we discuss the search for new physics with these
observables in more detail, and have also a closer look at the present
experimental situation.

\section{Discussion}\label{sec:disc}
Let us begin our discussion by turning first to the present experimental
situation. Concerning the direct CP asymmetries (\ref{def-AP}) and 
(\ref{def-A0}), we have
\begin{equation}\label{dir-exp}
{\cal A}_{\rm CP}^{{\rm dir}}=\left(26\pm19\right)\% 
\,\,\,\mbox{(BaBar \cite{babar})},\quad
{\cal A}_{\rm CP}^{(+)}=\left\{\begin{array}{ll}
(13\pm14)\% & \mbox{(BaBar \cite{babar})}\\
(-1.8\pm4.3\pm0.4)\% &\mbox{(CLEO \cite{cleo-dir}).}
\end{array}\right.
\end{equation}
The present status of the mixing-induced $B_d\to J/\psi K_{\rm S}$ 
CP asymmetry (\ref{def-mix}) is given as follows:\footnote{The values for
$\sin2\beta$ reported in \cite{babar,belle} are dominated by 
$B_d\to J/\psi K_{\rm S}$, but actually correspond to an average over 
various modes.}
\begin{equation}\label{B-factory}
-{\cal A}^{\mbox{{\scriptsize mix}}}_{\mbox{{\scriptsize CP}}}
=\left\{\begin{array}{ll}
0.79^{+0.41}_{-0.44}&\mbox{(CDF \cite{CDF})}\\
-0.10\pm0.42&\mbox{(BaBar \cite{babar})}\\
0.49^{+0.53}_{-0.57}&\mbox{(Belle \cite{belle}).}
\end{array}\right.
\end{equation}
Using the rates listed in \cite{PDG}, as well as $\tau_{B^+}/\tau_{B^0_d}=
1.060\pm0.029$, and adding the experimental errors in quadrature, we 
obtain\footnote{Phase-space effects play a negligible role in this 
expression.}
\begin{equation}\label{B-res}
B=\frac{\mbox{BR}(B_d\to J/\psi K)\tau_{B^+}/\tau_{B^0_d}-
\mbox{BR}(B^\pm\to J/\psi K^\pm)}{\mbox{BR}(B_d\to J/\psi K)
\tau_{B^+}/\tau_{B^0_d}+\mbox{BR}(B^\pm\to J/\psi K^\pm)}=(-2.9\pm 4.3)\%,
\end{equation}
where the numerical value depends rather sensitively on the lifetime ratio.
On the other hand, the BaBar results for the CP asymmetries listed in 
(\ref{dir-exp}) yield
\begin{equation}\label{SD-res}
S=(20\pm12)\%,\quad D=(7\pm12)\%.
\end{equation}

In view of the large experimental uncertainties, we cannot yet draw 
any conclusions. However, the situation should improve significantly
in the future. As can be seen in (\ref{S-expr})--(\ref{B-expr}), the
observable $S$ provides a ``smoking-gun'' signal of new-physics 
contributions to the $I=0$ sector, while $D$ and $B$ allow us to
probe new physics affecting the $I=1$ pieces. As the hierarchy in 
(\ref{ampl-hier1}) implies that $S$ receives terms of 
${\cal O}(\overline{\lambda})$, whereas $D$ and $B$ arise 
both at the $\overline{\lambda}^2$ level, we conclude that $S$ may
already be accessible at the first-generation $B$-factories (BaBar,
Belle, Tevatron-II), whereas the latter observables will probably 
be left for second-generation $B$ experiments (BTeV, LHCb). However,
should $B$ and $D$, in addition to $S$, also be found to be at the
$10\%$ level, i.e.\ should be measured at the first-generation
$B$-factories, we would not only have signals for physics
beyond the Standard Model, but also for large rescattering processes. 

A more pessimistic scenario one can imagine is that $S$ is measured at the 
$\overline{\lambda}^2$ level in the LHC era, whereas no indications for 
non-vanishing values of $D$ and $B$ are found. Then we would still have 
evidence for new physics, which would then correspond to 
$r_0^{(k)}={\cal O}(\overline{\lambda}^2)$ and 
$r_1^{(j)}={\cal O}(\overline{\lambda}^3)$. However, if all three 
observables are measured to be of ${\cal O}(\overline{\lambda}^2)$,
new-physics effects cannot be distinguished from Standard-Model 
contributions, which could also be enhanced to the 
$\overline{\lambda}^2$ level by large rescattering effects. This would be 
the most unfortunate case for the strategy to search for new-physics 
contributions to the $B\to J/\psi K$ decay amplitudes proposed in this 
paper. However, further information can be obtained with the help of 
the decay $B_s\to J/\psi K_{\rm S}$, which can be combined with 
$B_d\to J/\psi K_{\rm S}$ through the $U$-spin symmetry 
of strong interactions to extract the angle $\gamma$ of the unitarity 
triangle, and may shed light on new physics even in this case 
\cite{RF-BdspsiK}.  

As can be seen in (\ref{Amix-calc}), the mixing-induced CP asymmetry 
is affected both by $I=0$ and by $I=1$ new-physics contributions, where
the dominant ${\cal O}(\overline{\lambda})$ effects are expected to be
due to the $I=0$ sector. Neglecting terms of 
${\cal O}(\overline{\lambda}^2)$, we may write
\begin{equation}\label{Amix-simple}
{\cal A}^{\mbox{{\scriptsize mix}}}_{\mbox{{\scriptsize CP}}}=
-\sin(\phi+\delta\phi_{\rm NP}^{\rm dir}),
\end{equation}
with
\begin{equation}\label{PhiNP}
\delta\phi_{\rm NP}^{\rm dir}=
2\sum_k r_0^{(k)}\cos\delta_0^{(k)}\sin\varphi_0^{(k)}.
\end{equation}
The phase shift $\delta\phi_{\rm NP}^{\rm dir}={\cal  
O}(\overline{\lambda})$ 
may be as large as ${\cal O}(20^\circ)$. Since the Standard-Model range for 
$\phi$ is given by $28^\circ\leq\phi=2\beta\leq70^\circ$ \cite{AL}, the 
mixing-induced CP asymmetry (\ref{Amix-simple}) may also be affected 
significantly by new-physics contributions to the $B_d\to J/\psi K_{\rm S}$ 
decay amplitude, and not only in the ``standard'' fashion, through 
$B^0_d$--$\overline{B^0_d}$ mixing, as discussed in  
Subsection~\ref{sec:mix}. This would be another possibility to accommodate 
the small central value of the BaBar result in (\ref{B-factory}), which 
was not pointed out in \cite{sin2b-NP}. In order to gain confidence in such 
a scenario, it is crucial to improve also the measurements of the observables 
$S$, $D$ and $B$. Let us note that if, in addition to $S$, also $D$ and $B$ 
should be found to be sizeable, i.e.\ of ${\cal O}(\overline{\lambda})$, 
also the terms linear in $r_1^{(j)}$ have to be included in (\ref{PhiNP}):
\begin{equation}\label{PhiNP-res}
\left.\delta\phi_{\rm NP}^{\rm dir}\right|_{\rm res.}=
2\left[\sum_k r_0^{(k)}\cos\delta_0^{(k)}\sin\varphi_0^{(k)}+
\sum_j r_1^{(j)}\cos\delta_1^{(j)}\sin\varphi_1^{(j)}\right].
\end{equation}

So far, our considerations were completely general. Let us therefore  
comment briefly on a special case, where the strong phases $\delta_0^{(k)}$ 
and $\delta_1^{(j)}$ take the trivial values $0$ or $\pi$, as in 
``factorization''. In this case, (\ref{S-expr})--(\ref{B-expr}) 
would simplify as follows:
\begin{equation}
S\approx0,\quad D\approx0,\quad 
B\approx 2\sum_j r_1^{(j)}\sin\varphi_1^{(j)}={\cal  
O}(\overline{\lambda}^2),
\end{equation}
whereas (\ref{Amix-calc}) would yield
\begin{eqnarray}
{\cal A}^{\mbox{{\scriptsize mix}}}_{\mbox{{\scriptsize CP}}}&=&
-\sin\phi-2\,r_0\sin\varphi_0\cos\phi-2\,r_1\sin\varphi_1\cos\phi\nonumber\\
&&+r_0^2\Bigl[\left(1-\cos2\varphi_0\right)\sin\phi+
\sin2\varphi_0\cos\phi\Bigr]=-\sin\phi+
{\cal O}(\overline{\lambda}) + {\cal O}(\overline{\lambda}^2).
\end{eqnarray}
The important point is that $S$ and $D$ are governed by sines of the strong 
phases, whereas the new-physics contributions to $B$ and 
${\cal A}^{\rm mix}_{\rm CP}$ involve cosines of the corresponding 
strong phases. Consequently, these terms do {\it not} vanish for 
$\delta\to 0,\pi$. The impact of new physics on 
${\cal A}^{\rm mix}_{\rm CP}$ may still be sizeable in this scenario, 
whereas $B$ could only be measured in the LHC era \cite{LHC-Report}. If $S$ 
and $D$ should be observed at the $\overline{\lambda}$ and 
$\overline{\lambda}^2$ levels, respectively, we would not only get a 
``smoking-gun'' signal for new-physics contributions to the $B\to J/\psi K$ 
decay amplitudes, but also for non-factorizable hadronic effects. A 
measurement of all three observables $S$, $D$ and $B$ at the 
$\overline{\lambda}$ level would imply, in addition, large 
rescattering processes, as we have already emphasized above.

\section{Summary}\label{sec:concl}
We have presented a general analysis of new-physics effects in the 
$B^\pm\to J/\psi K^\pm$, $B_d\to J/\psi K_{\rm S}$ system. To this
end, we have taken into account the constraints that are implied by 
the $SU(2)$ isospin symmetry of strong interactions, and have estimated
the generic size of the new-physics contributions through dimensional
arguments following from the picture of effective field theory. In
addition to the usual mixing-induced CP asymmetry 
${\cal A}_{\rm CP}^{\rm mix}$ of the $B_d\to J/\psi K_{\rm S}$ mode, 
we have introduced a set of three observables, $S$, $D$ and $B$, which 
play the key role to search for ``smoking-gun'' signals of new-physics 
contributions to the $I=0$ and $I=1$ decay-amplitude sectors. Imposing
a plausible dynamical hierarchy of amplitudes, we argue that $S$ may
already be accessible at the first-generation $B$-factories, whereas 
the remaining ones will probably be left for second-generation $B$
experiments of the LHC era. However, in the presence of large rescattering 
effects, all three new-physics observables $S$, $D$ and $B$ may be sizeable.
At present, the large experimental uncertainties on these quantities do 
not allow us to draw any conclusions, and we strongly encourage our 
experimental colleagues to focus not only on ${\cal A}_{\rm CP}^{\rm mix}$, 
but also on $S$, $D$ and $B$. A future measurement corresponding to the 
central values given in (\ref{SD-res}), i.e.\ $S=20\%$ and $D=7\%$, would 
be as exciting as ${\cal A}_{\rm CP}^{\rm mix}=10\%$. Also the latter 
result could be due to new-physics contributions to the 
$B_d\to J/\psi K_{\rm S}$ decay amplitudes, and would not necessarily be
an indication of new physics in $B^0_d$--$\overline{B^0_d}$ mixing. We 
look forward to better data on $B\to J/\psi K$ decays, which will,  
hopefully, open a window to the physics beyond the Standard Model.

\section*{Acknowledgements}
The work of T.M. is supported by the DFG Graduiertenkolleg
``Elementarteilchenphysik an Beschleunigern'', by the
DFG Forschergruppe ``Quantenfeldtheorie, Computeralgebra und Monte  
Carlo Simulationen'', and by the
``Ministerium f\"ur Bildung und Forschung'' (bmb+f).

\end{document}